\documentclass[aps,prb,reprint,longbibliography,floatfix,fleqn]{revtex4-1}
\usepackage[utf8]{inputenc}
\usepackage{amsmath,graphicx,upgreek,amssymb,xcolor}
\usepackage[colorlinks=true,urlcolor=purple,citecolor=purple,filecolor=purple,linkcolor=purple]{hyperref}
\usepackage[english]{babel}

\makeatletter
\def\bbl@set@language#1{%
  \edef\languagename{%
    \ifnum\escapechar=\expandafter`\string#1\@empty
    \else\string#1\@empty\fi}%
  \@ifundefined{babel@language@alias@\languagename}{}{%
    \edef\languagename{\@nameuse{babel@language@alias@\languagename}}%
  }%
  \select@language{\languagename}%
  \expandafter\ifx\csname date\languagename\endcsname\relax\else
    \if@filesw
      \protected@write\@auxout{}{\string\select@language{\languagename}}%
      \bbl@for\bbl@tempa\BabelContentsFiles{%
        \addtocontents{\bbl@tempa}{\xstring\select@language{\languagename}}}%
      \bbl@usehooks{write}{}%
    \fi
  \fi}
\newcommand{\DeclareLanguageAlias}[2]{%
  \global\@namedef{babel@language@alias@#1}{#2}%
}
\makeatother

\DeclareLanguageAlias{en}{english}

\def\urusi{URu$_{\text2}$Si$_{\text2}$}

\def\ee{{\text{\textsc{elastic}}}} % "elastic"
\def\ii{{\text{\textsc{int}}}} % "interaction"

\def\Dfh{D$_{\text{4h}}$}

\def\Aog{{\text A_{\text{1g}}}}
\def\Atg{{\text A_{\text{2g}}}}
\def\Bog{{\text B_{\text{1g}}}}
\def\Btg{{\text B_{\text{2g}}}}
\def\Eg {{\text E_{\text  g}}}

\def\Atu{{\text A_{\text{2u}}}}

\def\Eu {{\text E_{\text  u}}}

\def\X{\text X}
\def\Y{\text Y}

\def\K{\text K}
\def\GPa{\text{GPa}}
\def\A{\text{\r A}}

\def\op{\textsc{op}} % order parameter
\def\ho{\textsc{ho}} % hidden order
\def\rus{\textsc{rus}} % resonant ultrasound spectroscopy 
\def\Rus{\textsc{Rus}} % Resonant ultrasound spectroscopy 
\def\afm{\textsc{afm}} % antiferromagnetism 
\def\recip{{\{-1\}}} % functional reciprocal

\begin{document}

\title{Elastic properties of hidden order in \urusi\ are reproduced by a staggered nematic}
\author{Jaron Kent-Dobias}
\author{Michael Matty}
\author{B.~J. Ramshaw}
\affiliation{
  Laboratory of Atomic \& Solid State Physics, Cornell University,
  Ithaca, NY, USA
}

\date\today

\begin{abstract}
  We develop a phenomenological mean field theory describing the hidden order
  phase in \urusi\ as a nematic of the $\Bog$ representation staggered along
  the $c$-axis. Several experimental features are reproduced by this theory:
  the topology of the temperature--pressure phase diagram, the response of the
  elastic modulus $(C_{11}-C_{12})/2$ above the transition at ambient pressure,
  and orthorhombic symmetry breaking in the high-pressure antiferromagnetic
  phase.  In this scenario, hidden order is characterized by broken rotational
  symmetry that is modulated along the $c$-axis, the primary order of the
  high-pressure phase is an unmodulated nematic, and the triple point
  joining those two phases with the high-temperature paramagnetic phase is a
  Lifshitz point.
\end{abstract}

\maketitle

\section{Introduction}

\urusi\ is a paradigmatic example of a material with an ordered state whose
broken symmetry remains unknown. This state, known as \emph{hidden order}
(\ho), sets the stage for unconventional superconductivity that emerges at even
lower temperatures.  At sufficiently large hydrostatic pressures, both
superconductivity and \ho\ give way to local moment antiferromagnetism
(\afm).\cite{Hassinger_2008}  Modern theories~\cite{Kambe_2018, Haule_2009,
Kusunose_2011_On, Kung_2015, Cricchio_2009, Ohkawa_1999, Santini_1994,
Kiss_2005, Harima_2010, Thalmeier_2011, Tonegawa_2012_Cyclotron,
Rau_2012_Hidden, Riggs_2015_Evidence, Hoshino_2013_Resolution,
Ikeda_1998_Theory, Chandra_2013_Hastatic, Harrison_2019_Hidden, Ikeda_2012} propose
associating any of a variety of broken symmetries with \ho. Motivated by the anomalous temperature dependence of one of the elastic moduli , this work analyzes
a family of phenomenological models with order parameters of general symmetry
that couple linearly to strain. Of these, only one is compatible with two
experimental observations: first, the $\Bog$ ``nematic" elastic susceptibility
$(C_{11}-C_{12})/2$ softens anomalously from room temperature down to
$T_{\text{\ho}}=17.5\,\K$;\cite{deVisser_1986_Thermal} and second, a $\Bog$ nematic
distortion is observed by x-ray scattering under sufficient pressure to destroy
the \ho\ state.\cite{Choi_2018}

Recent resonant ultrasound spectroscopy (\rus) measurements were used to
examine the thermodynamic discontinuities in the elastic moduli at
$T_{\text{\ho}}$.\cite{Ghosh_2020_One-component} The observation of
discontinuities only in compressional, or $\Aog$, elastic moduli requires that
the point-group representation of \ho\ be one-dimensional. This rules out many
order parameter candidates~\cite{Thalmeier_2011, Tonegawa_2012_Cyclotron,
Rau_2012_Hidden, Riggs_2015_Evidence, Hoshino_2013_Resolution, Ikeda_2012,
Chandra_2013_Origin} in a model-independent way, but doesn't differentiate
between those that remain. 

Recent x-ray experiments discovered rotational symmetry breaking in \urusi\
under pressure.\cite{Choi_2018} Above 0.13--0.5 $\GPa$ (depending on
temperature), \urusi\ undergoes a $\Bog$ nematic distortion, which might be
related to the anomalous softening of the $\Bog$ elastic modulus
$(C_{11}-C_{12})/2$ that occurs over a broad temperature range at zero
pressure.\cite{Wolf_1994, Kuwahara_1997,yanagisawa2012gamma3} Motivated by these results---which
hint at a $\Bog$ strain susceptibility associated with the \ho\ state---we
construct a phenomenological mean field theory for an arbitrary \op\ coupled to
strain, and then determine the effect of its phase transitions on the elastic
response in different symmetry channels.

We find that only one \op\ representation reproduces the anomalous $\Bog$
elastic modulus, which softens in a Curie--Weiss-like manner from room
temperature and then cusps at $T_{\text{\ho}}$. That theory associates \ho\
with a $\Bog$ \op\ modulated along the $c$-axis, the high pressure state with
uniform $\Bog$ order, and the triple point between them with a Lifshitz point.
In addition to the agreement with the ultrasound data across a broad
temperature range, our model predicts uniform $\Bog$ strain at high
pressure---the same distortion that was recently seen in x-ray scattering
experiments.\cite{Choi_2018} This work strongly motivates future ultrasound
experiments under pressure approaching the Lifshitz point, which should find
that the $(C_{11}-C_{12})/2$ modulus diverges as the uniform $\Bog$ strain of
the high pressure phase is approached.

\section{Model \& Phase Diagram}

The point group of \urusi\ is \Dfh, and any theory must locally respect this
symmetry in the high-temperature phase. Our phenomenological free energy
density contains three parts: the elastic free energy, the \op, and the
interaction between strain and \op. The most general quadratic free energy of
the strain $\epsilon$ is $f_\ee=C^0_{ijkl}\epsilon_{ij}\epsilon_{kl}$.
\footnote{Components of the elastic modulus tensor $C$ were given in the
  popular Voigt notation in the abstract and introduction. Here and henceforth
the notation used is that natural for a rank-four tensor.} The form of the bare
moduli tensor $C^0$ is further restricted by symmetry.
\cite{Landau_1986_Theory} Linear combinations of the six independent components
of strain form five irreducible components of strain in \Dfh\ as
\begin{equation}
  \begin{aligned}
    & \epsilon_{\Aog,1}=\epsilon_{11}+\epsilon_{22} \hspace{0.15\columnwidth} &&
      \epsilon_\Bog=\epsilon_{11}-\epsilon_{22} \\
    & \epsilon_{\Aog,2}=\epsilon_{33} &&
      \epsilon_\Btg=2\epsilon_{12} \\
    & \epsilon_\Eg=2\{\epsilon_{11},\epsilon_{22}\}.
  \end{aligned}
  \label{eq:strain-components}
\end{equation}
All quadratic combinations of these irreducible strains that transform like
$\Aog$ are included in the free energy,
\begin{equation}
  f_\ee=\frac12\sum_\X C^0_{\X,ij}\epsilon_{\X,i}\epsilon_{\X,j},
\end{equation}
where the sum is over irreducible representations of the point group and the
bare elastic moduli $C^0_\X$ are 
\begin{equation}
  \begin{aligned}
    & C^0_{\Aog,11}=\tfrac12(C^0_{1111}+C^0_{1122}) &&
      C^0_{\Bog}=\tfrac12(C^0_{1111}-C^0_{1122}) \\
    & C^0_{\Aog,22}=C^0_{3333} &&
      C^0_{\Btg}=C^0_{1212} \\
    & C^0_{\Aog,12}=C^0_{1133} &&
      C^0_{\Eg}=C^0_{1313}.
  \end{aligned}
\end{equation}
The interaction between strain and an \op\ $\eta$ depends on the point group
representation of $\eta$. If this representation is $\X$, the most general
coupling to linear order is
\begin{equation}
  f_\ii=-b^{(i)}\epsilon_\X^{(i)}\eta.
\end{equation}
Many high-order interations are permitted, and in the appendix another of the
form $\epsilon^2\eta^2$ is added to the following analysis.
If there exists no component of strain that transforms like the representation
$\X$ then there can be no linear coupling. The next-order coupling is linear in
strain, quadratic in order parameter, and the effect of this coupling at a
continuous phase transition is to produce a jump in the $\Aog$ elastic moduli
if $\eta$ is single-component, \cite{Luthi_1970, Ramshaw_2015, Shekhter_2013}
and jumps in other elastic moduli if multicomponent.\cite{Ghosh_2020_One-component} Because
we are interested in physics that anticipates the phase transition---for
instance, that the growing \op\ susceptibility is reflected directly in the
elastic susceptibility---we will focus our attention on \op s that can produce
linear couplings to strain.  Looking at the components present in
\eqref{eq:strain-components}, this rules out all of the u-reps (which are odd
under inversion), the $\Atg$ irrep, and all half-integer (spinor)
representations.

If the \op\ transforms like $\Aog$ (e.g. a fluctuation in valence number), odd
terms are allowed in its free energy and without fine-tuning any transition
will be first order and not continuous. Since the \ho\ phase transition is
second-order,\cite{deVisser_1986_Thermal} we will henceforth rule out $\Aog$ \op s as
well.  For the \op\ representation $\X$ as any of those remaining---$\Bog$,
$\Btg$, or $\Eg$---the most general quadratic free energy density is
\begin{equation}
  \begin{aligned}
    f_\op=\frac12\big[&r\eta^2+c_\parallel(\nabla_\parallel\eta)^2
      +c_\perp(\nabla_\perp\eta)^2 \\
      &\qquad\qquad\qquad\quad+D_\perp(\nabla_\perp^2\eta)^2\big]+u\eta^4,
  \end{aligned}
  \label{eq:fo}
\end{equation}
where $\nabla_\parallel=\{\partial_1,\partial_2\}$ transforms like $\Eu$, and
$\nabla_\perp=\partial_3$ transforms like $\Atu$. Other quartic terms are
allowed---especially many for an $\Eg$ \op---but we have included only those
terms necessary for stability when either $r$ or $c_\perp$ become negative as a function of temperature. The
full free energy functional of $\eta$ and $\epsilon$ is
\begin{equation}
  \begin{aligned}
    F[\eta,\epsilon]
      &=F_\op[\eta]+F_\ee[\epsilon]+F_\ii[\eta,\epsilon] \\
      &=\int dx\,(f_\op+f_\ee+f_\ii).
  \end{aligned}
  \label{eq:free_energy}
\end{equation}

Rather than analyze this two-argument functional directly, we begin by tracing
out the strain and studying the behavior of the \op\ alone. Later we will invert
this procedure and trace out the \op\ when we compute the effective elastic
moduli. The only strain relevant to an \op\ of representation $\X$ at linear coupling is
$\epsilon_\X$, which can be traced out of the problem exactly in mean field
theory. Extremizing the functional \eqref{eq:free_energy} with respect to
$\epsilon_\X$ gives
\begin{equation}
  0
    =\frac{\delta F[\eta,\epsilon]}{\delta\epsilon_\X(x)}\bigg|_{\epsilon=\epsilon_\star}
    =C^0_\X\epsilon^\star_\X(x)-b\eta(x),
\end{equation}
which in turn gives the strain field conditioned on the state of the \op\ field
as $\epsilon_\X^\star[\eta](x)=(b/C^0_\X)\eta(x)$ at all spatial coordinates
$x$, and $\epsilon_\Y^\star[\eta]=0$ for all other irreps $\Y\neq\X$. Upon
substitution into \eqref{eq:free_energy}, the resulting single-argument
free energy functional $F[\eta,\epsilon_\star[\eta]]$ has a density identical
to $f_\op$ with the identification $r\to\tilde r=r-b^2/2C^0_\X$. 

\begin{figure}[htpb]
  \includegraphics[width=\columnwidth]{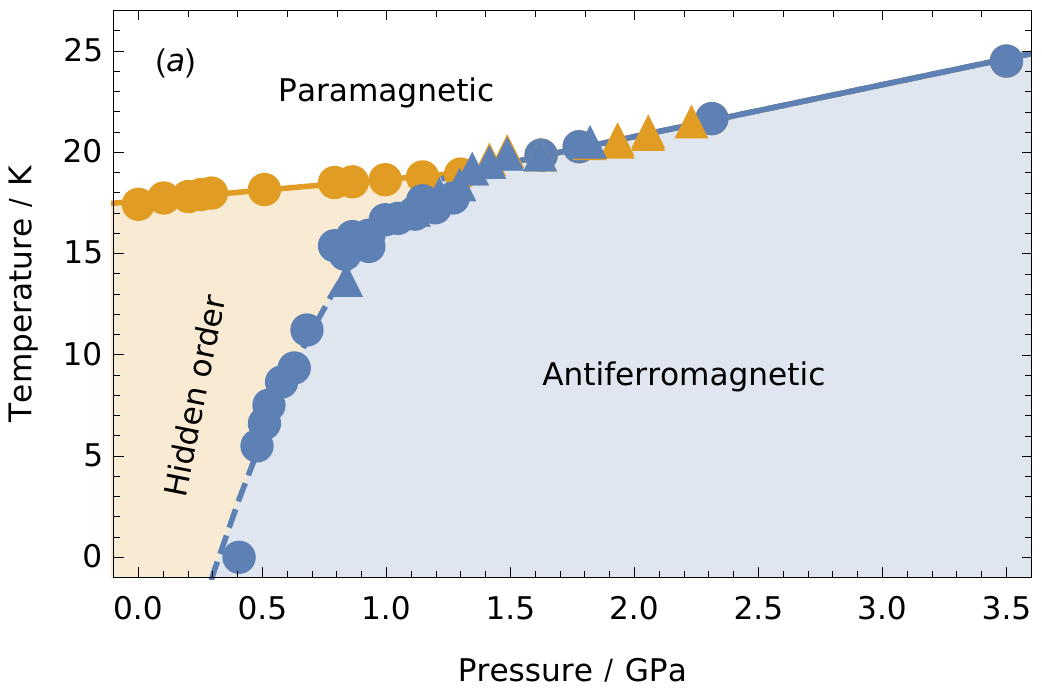}

  \vspace{1em}

  \includegraphics[width=0.51\columnwidth]{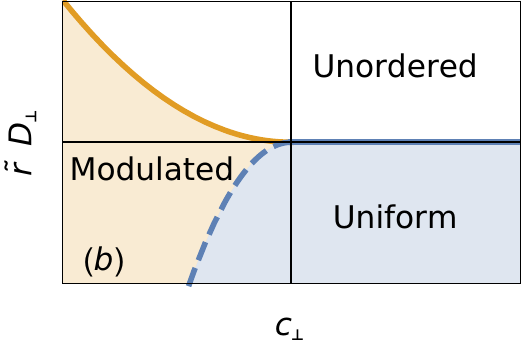}\hspace{-1.5em}
  \includegraphics[width=0.51\columnwidth]{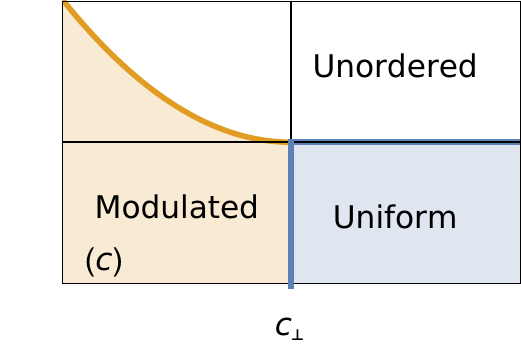}
  \caption{
    Phase diagrams for (a) \urusi\ from experiments (neglecting the
    superconducting phase)~\cite{Hassinger_2008} (b) mean field theory of a
    one-component ($\Bog$ or $\Btg$) Lifshitz point (c) mean field theory of a
    two-component ($\Eg$) Lifshitz point. Solid lines denote second-order
    transitions, while dashed lines denote first order transitions.  Later,
    when we fit the elastic moduli predictions for a $\Bog$ \op\ to data along
    the ambient pressure line, we will take $\Delta\tilde r=\tilde r-\tilde
    r_c=a(T-T_c)$.
  }
  \label{fig:phases}
\end{figure}

With the strain traced out, \eqref{eq:fo} describes the theory of a Lifshitz
point at $\tilde r=c_\perp=0$.\cite{Lifshitz_1942_OnI, Lifshitz_1942_OnII} The
properties discussed in the remainder of this section can all be found in a
standard text, e.g., in Chapter 4 \S6.5  of Chaikin \&
Lubensky.\cite{Chaikin_1995} For a one-component \op\ ($\Bog$ or $\Btg$) and
positive $c_\parallel$, it is traditional to make the field ansatz
$\langle\eta(x)\rangle=\eta_*\cos(q_*x_3)$. For $\tilde r>0$ and $c_\perp>0$,
or $\tilde r>c_\perp^2/4D_\perp$ and $c_\perp<0$, the only stable solution is
$\eta_*=q_*=0$ and the system is unordered. For $\tilde r<0$ there are free
energy minima for $q_*=0$ and $\eta_*^2=-\tilde r/4u$ and this system has
uniform order of the \op\ representation, e.g., $\Bog$ or $\Btg$. For
$c_\perp<0$ and $\tilde r<c_\perp^2/4D_\perp$ there are free energy minima for
$q_*^2=-c_\perp/2D_\perp$ and
\begin{equation}
  \eta_*^2=\frac{c_\perp^2-4D_\perp\tilde r}{12D_\perp u}
    =\frac{\tilde r_c-\tilde r}{3u}
    =\frac{|\Delta\tilde r|}{3u},
\end{equation}
with $\tilde r_c=c_\perp^2/4D_\perp$ and the system has modulated order. The
transition between the uniform and modulated orderings is first order for a
one-component \op\ and occurs along the line $c_\perp=-2\sqrt{-D_\perp\tilde
r/5}$.

For a two-component \op\ ($\Eg$) we must also allow a relative phase between
the two components of the \op. In this case the uniform ordered phase is only
stable for $c_\perp>0$, and the modulated phase is now characterized by helical
order with $\langle\eta(x)\rangle=\eta_*\{\cos(q_*x_3),\sin(q_*x_3)\}$.  The
uniform to modulated transition is now continuous. This does not reproduce the
physics of \urusi, whose \ho\ phase is bounded by a line of first order
transitions at high pressure, and so we will henceforth neglect the possibility
of a multicomponent order parameter---consistent with earlier ultrasound measurements \cite{Ghosh_2020_One-component}. Schematic phase diagrams for both the one-
and two-component models are shown in Figure~\ref{fig:phases}.

\section{Susceptibility \& Elastic Moduli}
We will now derive the effective elastic tensor $C$ that results from the
coupling of strain to the \op. The ultimate result, found in
\eqref{eq:elastic.susceptibility}, is that $C_\X$ differs from its bare value
$C^0_\X$ only for the representation $\X$ of the \op. Moreover, this modulus
does not vanish at the unordered to modulated transition---as it would if the
transition were a $q=0$ phase transition---but instead ends in a cusp. In this
section we start by computing the susceptibility of the \op\ at the unordered
to modulated transition, and then compute the elastic modulus for the same. 

The susceptibility of a single-component ($\Bog$ or $\Btg$) \op\ is
\begin{equation}
  \begin{aligned}
    &\chi^\recip(x,x')
      =\frac{\delta^2F[\eta,\epsilon_\star[\eta]]}{\delta\eta(x)\delta\eta(x')}\bigg|_{\eta=\langle\eta\rangle}
      =\big[\tilde r-c_\parallel\nabla_\parallel^2 \\
        &\qquad-c_\perp\nabla_\perp^2+D_\perp\nabla_\perp^4+12u\langle\eta(x)\rangle^2\big]\delta(x-x'),
  \end{aligned}
  \label{eq:sus_def}
\end{equation}
where $\recip$ indicates a functional reciprocal defined as
\begin{equation}
  \int dx''\,\chi^\recip(x,x'')\chi(x'',x')=\delta(x-x').
\end{equation}
Taking the Fourier transform and integrating out $q'$ gives
\begin{equation}
  \chi(q)
    =\big(\tilde r+c_\parallel q_\parallel^2+c_\perp q_\perp^2+D_\perp q_\perp^4
    +12u\sum_{q'}\langle\tilde\eta_{q'}\rangle\langle\tilde\eta_{-q'}\rangle\big)^{-1}.
\end{equation}
Near the unordered to modulated transition this yields
\begin{equation}
  \begin{aligned}
    \chi(q)
      &=\big[c_\parallel q_\parallel^2+D_\perp(q_*^2-q_\perp^2)^2
    +|\Delta\tilde r|\big]^{-1} \\
      &=\frac1{D_\perp}\frac{\xi_\perp^4}
        {1+\xi_\parallel^2q_\parallel^2+\xi_\perp^4(q_*^2-q_\perp^2)^2},
  \end{aligned}
  \label{eq:susceptibility}
\end{equation}
with $\xi_\perp=(|\Delta\tilde r|/D_\perp)^{-1/4}=\xi_{\perp0}|t|^{-1/4}$ and
$\xi_\parallel=(|\Delta\tilde
r|/c_\parallel)^{-1/2}=\xi_{\parallel0}|t|^{-1/2}$, where $t=(T-T_c)/T_c$ is
the reduced temperature and $\xi_{\perp0}=(D_\perp/aT_c)^{1/4}$ and
$\xi_{\parallel0}=(c_\parallel/aT_c)^{1/2}$ are the bare correlation lengths
perpendicular and parallel to the plane, respectively.  The static
susceptibility $\chi(0)=(D_\perp q_*^4+|\Delta\tilde r|)^{-1}$ does not diverge
at the unordered to modulated transition. Though it anticipates a transition
with Curie--Weiss-like divergence at the lower point $a(T-T_c)=\Delta\tilde
r=-D_\perp q_*^4<0$, this is cut off with a cusp at the phase transition at $\Delta\tilde r=0$. 

The elastic susceptibility, which is the reciprocal of the effective elastic
modulus, is found in a similar way to the \op\ susceptibility: we must trace
over $\eta$ and take the second variation of the resulting effective free
energy functional of $\epsilon$ alone. Extremizing over $\eta$ yields
\begin{equation}
  0=\frac{\delta F[\eta,\epsilon]}{\delta\eta(x)}\bigg|_{\eta=\eta_\star}
    =\frac{\delta F_\op[\eta]}{\delta\eta(x)}\bigg|_{\eta=\eta_\star}-b\epsilon_\X(x),
  \label{eq:implicit.eta}
\end{equation}
which implicitly gives $\eta_\star[\epsilon]$, the \op\ conditioned
on the configuration of the strain. Since $\eta_\star$ is a functional of $\epsilon_\X$
alone, only the modulus $C_\X$ will be modified from its bare value $C^0_\X$.

Though the differential equation for $\eta_\star$ cannot be solved explicitly, we
can use the inverse function theorem to make use of \eqref{eq:implicit.eta} anyway.
First, denote by $\eta_\star^{-1}[\eta]$ the inverse functional of $\eta_\star$
implied by \eqref{eq:implicit.eta}, which gives the function $\epsilon_\X$
corresponding to each solution of \eqref{eq:implicit.eta} it receives. This we
can immediately identify from \eqref{eq:implicit.eta} as
$\eta^{-1}_\star[\eta](x)=b^{-1}(\delta F_\op[\eta]/\delta\eta(x))$.  Now, we
use the inverse function theorem to relate the functional reciprocal of the
derivative of $\eta_\star[\epsilon]$ with respect to $\epsilon_\X$ to the
derivative of $\eta^{-1}_\star[\eta]$ with respect to $\eta$, yielding
\begin{equation}
  \begin{aligned}
    \bigg(\frac{\delta\eta_\star[\epsilon](x)}{\delta\epsilon_\X(x')}\bigg)^\recip
    &=\frac{\delta\eta_\star^{-1}[\eta](x)}{\delta\eta(x')}\bigg|_{\eta=\eta_\star[\epsilon]} \\
    &=b^{-1}\frac{\delta^2F_\op[\eta]}{\delta\eta(x)\delta\eta(x')}\bigg|_{\eta=\eta_\star[\epsilon]}.
  \end{aligned}
  \label{eq:inv.func}
\end{equation}
Next, \eqref{eq:implicit.eta} and \eqref{eq:inv.func} can be used in concert
with the ordinary rules of functional calculus to yield the second variation
\begin{widetext}
\begin{equation}
  \begin{aligned}
    &\frac{\delta^2F[\eta_\star[\epsilon],\epsilon]}{\delta\epsilon_\X(x)\delta\epsilon_\X(x')} 
    =C^0_\X\delta(x-x')-
    2b\frac{\delta\eta_\star[\epsilon](x)}{\delta\epsilon_\X(x')}
    -b\int dx''\,\frac{\delta^2\eta_\star[\epsilon](x)}{\delta\epsilon_\X(x')\delta\epsilon_\X(x'')}\epsilon_\X(x'') \\
    &\qquad\qquad\qquad+\int dx''\,\frac{\delta^2\eta_\star[\epsilon](x'')}{\delta\epsilon_\X(x)\delta\epsilon_\X(x')}\frac{\delta F_\op[\eta]}{\delta\eta(x'')}\bigg|_{\eta=\eta_\star[\epsilon]}
    +\int dx''\,dx'''\,\frac{\delta\eta_\star[\epsilon](x'')}{\delta\epsilon_\X(x)}\frac{\delta\eta_\star[\epsilon](x''')}{\delta\epsilon_\X(x')}\frac{\delta^2F_\op[\eta]}{\delta\eta(x'')\delta\eta(x''')}\bigg|_{\eta=\eta_\star[\epsilon]} \\ 
    &\qquad=C^0_\X\delta(x-x')-
    2b\frac{\delta\eta_\star[\epsilon](x)}{\delta\epsilon_\X(x')}
    -b\int dx''\,\frac{\delta^2\eta_\star[\epsilon](x)}{\delta\epsilon_\X(x')\delta\epsilon_\X(x'')}\epsilon_\X(x'') \\
    &\qquad\qquad\qquad\qquad+\int dx''\,\frac{\delta^2\eta_\star[\epsilon](x'')}{\delta\epsilon_\X(x)\delta\epsilon_\X(x')}(b\epsilon_\X(x''))
    +b\int dx''\,dx'''\,\frac{\delta\eta_\star[\epsilon](x'')}{\delta\epsilon_\X(x)}\frac{\delta\eta_\star[\epsilon](x''')}{\delta\epsilon_\X(x')} \bigg(\frac{\partial\eta_\star[\epsilon](x'')}{\partial\epsilon_\X(x''')}\bigg)^\recip\\ 
    &\qquad=C^0_\X\delta(x-x')-
    2b\frac{\delta\eta_\star[\epsilon](x)}{\delta\epsilon_\X(x')}
    +b\int dx''\,\delta(x-x'')\frac{\delta\eta_\star[\epsilon](x'')}{\delta\epsilon_\X(x')} 
    =C^0_\X\delta(x-x')-b\frac{\delta\eta_\star[\epsilon](x)}{\delta\epsilon_\X(x')}.
  \end{aligned}
  \label{eq:big.boy}
\end{equation}
\end{widetext}
The elastic modulus is given by the second variation \eqref{eq:big.boy}
evaluated at the extremized strain $\langle\epsilon\rangle$. To calculate it,
note that evaluating the second variation of $F_\op$ in \eqref{eq:inv.func} at
$\langle\epsilon\rangle$ (or
$\eta_\star(\langle\epsilon\rangle)=\langle\eta\rangle$) yields
\begin{equation}
  \bigg(\frac{\delta\eta_\star[\epsilon](x)}{\delta\epsilon_\X(x')}\bigg)^\recip\bigg|_{\epsilon=\langle\epsilon\rangle}
    =b^{-1}\chi^\recip(x,x')+\frac{b}{C^0_\X}\delta(x-x'),
  \label{eq:recip.deriv.op}
\end{equation}
where $\chi^\recip$ is the \op\ susceptibility given by \eqref{eq:sus_def}.
Upon substitution into \eqref{eq:big.boy} and taking the Fourier transform of
the result, we finally arrive at
\begin{equation}
  C_\X(q)
  =C^0_\X-b\bigg(\frac1{b\chi(q)}+\frac b{C^0_\X}\bigg)^{-1}
  =C^0_\X\bigg(1+\frac{b^2}{C^0_\X}\chi(q)\bigg)^{-1}.
  \label{eq:elastic.susceptibility}
\end{equation}
Though not relevant here, this result generalizes to multicomponent \op s.

What does \eqref{eq:elastic.susceptibility} predict in the vicinity of the \ho\
transition? Near the disordered to modulated transition---the zero-pressure
transition to the HO state---the static modulus is given by
\begin{equation}
  C_\X(0)=C_\X^0\bigg[1+\frac{b^2}{C_\X^0}\big(D_\perp q_*^4+|\Delta\tilde r|\big)^{-1}\bigg]^{-1}.
  \label{eq:static_modulus}
\end{equation}
This corresponds to a softening in the $\X$-modulus approaching the transition
that is cut off with a cusp of the form $|\Delta\tilde
r|^\gamma\propto|T-T_c|^\gamma$ with $\gamma=1$. This is our main result. The
only \op\ irreps that couple linearly with strain and reproduce the topology of
the \urusi\ phase diagram are $\Bog$ and $\Btg$. For either of these irreps,
the transition into a modulated rather than uniform phase masks traditional
signatures of a continuous transition by locating thermodynamic singularities
at nonzero $q=q_*$.  The remaining clue at $q=0$ is a particular kink in the
corresponding modulus.

\section{Comparison to experiment}

\begin{figure*}[htpb]
  \centering
  \includegraphics{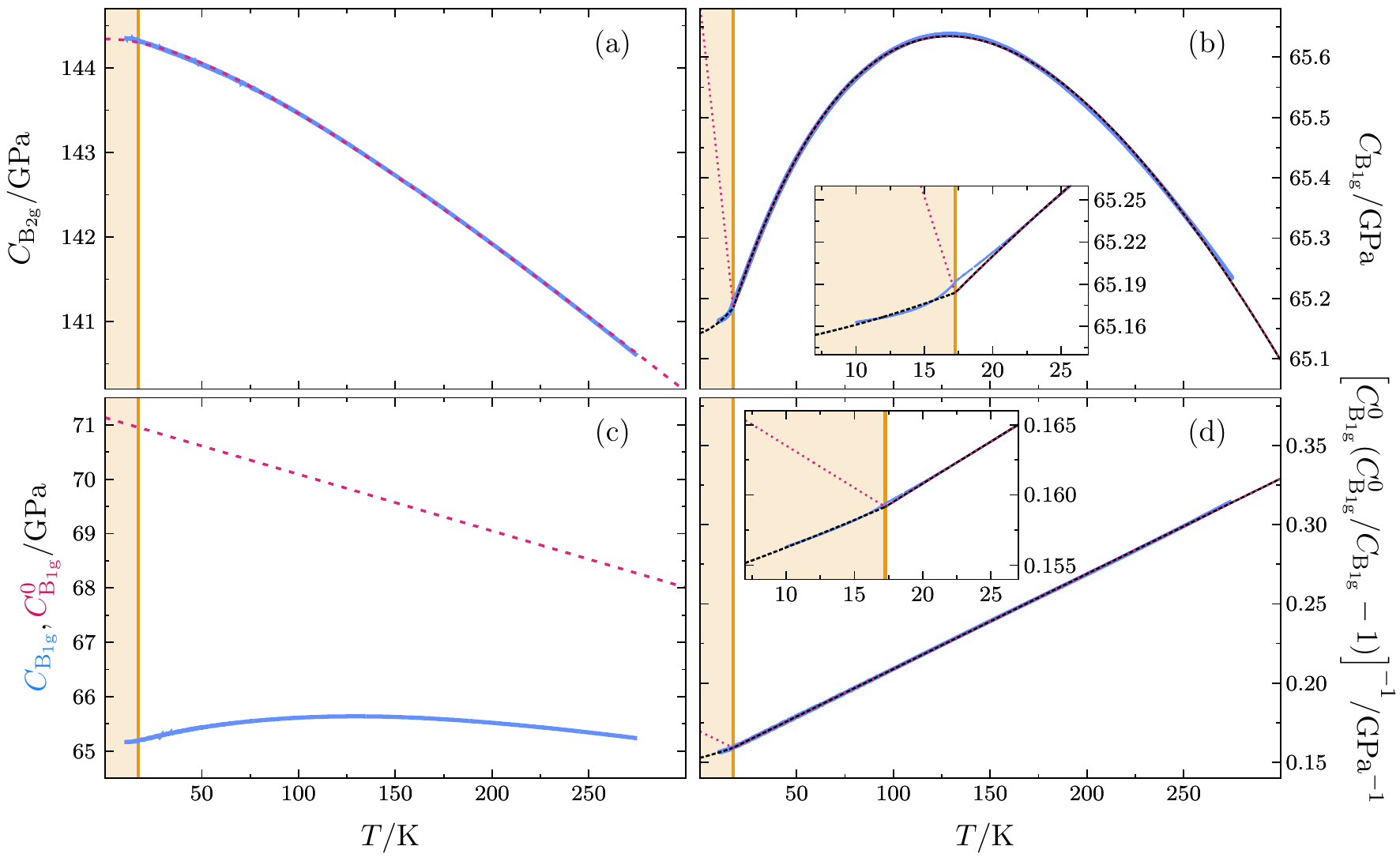}
  \caption{
    \Rus\ measurements of the elastic moduli of \urusi\ at ambient pressure as
    a function of temperature from recent
    experiments\cite{Ghosh_2020_One-component} (blue, solid) alongside fits to
    theory (magenta, dashed and black, dashed). The solid yellow region shows
    the location of the \ho\ phase. (a) $\Btg$ modulus data and a fit to the
    standard form.\cite{Varshni_1970} (b) $\Bog$ modulus data and a fit to
    \eqref{eq:static_modulus} (magenta, dashed) and a fit to \eqref{eq:C0}
    (black, dashed). The fit gives
    $C^0_\Bog\simeq\big[71-(0.010\,\K^{-1})T\big]\,\GPa$, $b^2/D_\perp
    q_*^4\simeq6.28\,\GPa$, and $b^2/a\simeq1665\,\GPa\,\K^{-1}$. Addition of a
    quadratic term in $C^0_\Bog$ was here not needed for the
    fit.\cite{Varshni_1970} (c) $\Bog$ modulus data and the fit of the
    \emph{bare} $\Bog$ modulus. (d) $\Bog$ modulus data and the fits
    transformed by $[C^0_\Bog(C^0_\Bog/C_\Bog-1)]]^{-1}$, which is predicted
    from \eqref{eq:static_modulus} to equal $D_\perp q_*^4/b^2+a/b^2|T-T_c|$,
    e.g., an absolute value function.
  }
  \label{fig:data}
\end{figure*}

\Rus\ experiments~\cite{Ghosh_2020_One-component} yield the individual elastic moduli
broken into irreps; data for the $\Bog$ and $\Btg$ components defined in
\eqref{eq:strain-components} are shown in Figures \ref{fig:data}(a--b).  The
$\Btg$ modulus in Fig.~\ref{fig:data}(a) doesn't appear to have any response to
the presence of the transition, exhibiting the expected linear stiffening upon
cooling from room temperature, with a low-temperature cutoff at some fraction
of the Debye temperature.\cite{Varshni_1970} The $\Bog$ modulus
Fig.~\ref{fig:data}(b) has a dramatic response, softening over the course of
roughly $100\,\K$ and then cusping at the \ho\ transition. The data in the
high-temperature phase can be fit to the theory \eqref{eq:static_modulus}, with
a linear background modulus $C^0_\Bog$ and $\tilde r-\tilde r_c=a(T-T_c)$, and
the result is shown in Figure \ref{fig:data}(b).

The behavior of the modulus below the transition does not match
\eqref{eq:static_modulus} well, but this is because of the truncation of the
free energy expansion used above.  Higher order terms like $\eta^2\epsilon^2$
and $\epsilon^4$ contribute to the modulus starting at order $\eta_*^2$ and
therefore change the behavior below the transition, where the expectation value of $\eta$ is finite, but not above it, where the expectation value of $\eta$ is zero. To
demonstrate this, in Appendix~\ref{sec:higher-order} we compute the modulus in
a theory where the interaction free energy is truncated after fourth order with
new term $\frac12g\eta^2\epsilon^2$. The dashed black line in
Fig.~\ref{fig:data} shows the fit of the \rus\ data to \eqref{eq:C0} and shows
that successive high-order corrections can account for the low-temperature
behavior. 

The data and theory appear quantitatively consistent, suggesting that \ho\ can
be described as a $\Bog$-nematic phase that is modulated at finite $q$ along
the $c-$axis. The predicted softening appears over hundreds of Kelvin; Figures
\ref{fig:data}(c--d) show the background modulus $C_\Bog^0$ and the
\op--induced response isolated from each other. 

We have seen that the mean-field theory of a $\Bog$ \op\ recreates the topology
of the \ho\ phase diagram and the temperature dependence of the $\Bog$ elastic
modulus at zero pressure. This theory has several other physical implications.
First, the association of a modulated $\Bog$ order with the \ho\ phase implies
a \emph{uniform} $\Bog$ order associated with the high pressure phase, and
moreover a uniform $\Bog$ strain of magnitude
$\langle\epsilon_\Bog\rangle^2=b^2\tilde r/4u(C^0_\Bog)^2$, which corresponds
to an orthorhombic structural phase.  The onset of orthorhombic symmetry
breaking was recently detected at high pressure in \urusi\ using x-ray
diffraction, a further consistency of this theory with the phenomenology of
\urusi.\cite{Choi_2018} 

Second, as the Lifshitz point is approached from low pressure, this theory
predicts that the modulation wavevector $q_*$ should vanish continuously. Far
from the Lifshitz point we expect the wavevector to lock into values
commensurate with the space group of the lattice, and moreover that at zero
pressure, where the \rus\ data here was collected, the half-wavelength of the
modulation should be commensurate with the lattice spacing $a_3\simeq9.68\,\A$,
or $q_*=\pi/a_3\simeq0.328\,\A^{-1}$.\cite{Bareille_2014_Momentum-resolved,
Yoshida_2010_Signature, Yoshida_2013_Translational, Meng_2013_Imaging,
Broholm_1991, Wiebe_2007, Bourdarot_2010, Hassinger_2010} In between these two
regimes, mean field theory predicts that the ordering wavevector shrinks by
jumping between ever-closer commensurate values in the style of the devil's
staircase.\cite{Bak_1982} In reality the presence of fluctuations may wash out
these transitions.

This motivates future ultrasound experiments done under pressure, where the
depth of the cusp in the $\Bog$ modulus should deepen (perhaps with these
commensurability jumps) at low pressure and approach zero as
$q_*^4\sim(c_\perp/2D_\perp)^2$ near the Lifshitz point.  Alternatively, \rus\
done at ambient pressure might examine the heavy Fermi liquid to \afm\
transition by doping. Though previous \rus\ studies have doped \urusi\ with
rhodium,\cite{Yanagisawa_2014} rhodium changes the carrier concentration as well as the lattice spacing, and may favour the promotion of the magnetic phase. An iso-electronic (as well as iso-magnetic) dopant such as iron may more faithfully explore the transition out of the HO phase. Our work also motivates experiments
that can probe the entire correlation function---like x-ray and neutron
scattering---and directly resolve its finite-$q$ divergence.  The presence of
spatial commensurability is known to be irrelevant to critical behavior at a
one-component disordered to modulated transition, and therefore is not expected
to modify the thermodynamic behavior otherwise.\cite{Garel_1976} 

There are two apparent discrepancies between the orthorhombic strain in the
phase diagram presented by recent x-ray data\cite{Choi_2018}, and that
predicted by our mean field theory if its uniform $\Bog$ phase is taken to be
coincident with \urusi's \afm.  The first is the apparent onset of the
orthorhombic phase in the \ho\ state at slightly lower pressures than the onset
of \afm.  As the recent x-ray research\cite{Choi_2018} notes, this misalignment
of the two transitions as function of doping could be due to the lack of an
ambient pressure calibration for the lattice constant. The second discrepancy
is the onset of orthorhombicity at higher temperatures than the onset of \afm.
We note that magnetic susceptibility data sees no trace of another phase
transition at these higher temperatures. \cite{Inoue_2001} It is therefore
possible that the high-temperature orthorhombic signature in x-ray scattering
is not the result of a bulk thermodynamic phase, but instead marks the onset of
short-range correlations, as it does in the high-T$_{\mathrm{c}}$ cuprates
\cite{Ghiringhelli_2012} (where the onset of CDW correlations also lacks a
thermodynamic phase transition). 

Three dimensions is below the upper critical dimension $4\frac12$ of a
one-component disordered-to-modulated transition, and so mean field theory
should break down sufficiently close to the critical point due to fluctuations,
at the Ginzburg temperature. \cite{Hornreich_1980, Ginzburg_1961_Some} Magnetic
phase transitions tend to have a Ginzburg temperature of order one.  Our fit
above gives $\xi_{\perp0}q_*=(D_\perp q_*^4/aT_c)^{1/4}\simeq2$, which combined
with the speculation of $q_*\simeq\pi/a_3$ puts the bare correlation length
$\xi_{\perp0}$ on the order of lattice constant, which is about what one would
expect for a generic magnetic transition.  The agreement of this data in the
$(T-T_{\rm{HO}})/T_{\rm{HO}}\sim0.1$--10 range with the mean field exponent suggests that this region is
outside the Ginzburg region, but an experiment may begin to see deviations from
mean field behavior within approximately several Kelvin of the critical point.
An ultrasound experiment with finer temperature resolution near the
critical point may be able to resolve a modified cusp exponent
$\gamma\simeq1.31$,\cite{Guida_1998_Critical} since according to one analysis
the universality class of a uniaxial modulated one-component \op\ is that of
the $\mathrm O(2)$, 3D XY transition.\cite{Garel_1976} A crossover from mean
field theory may explain the small discrepancy in our fit very close to the
critical point.

\section{Conclusion and Outlook.} We have developed a general phenomenological
treatment of  \ho\ \op s that have the potential for linear coupling to strain.
The two representations with mean field phase diagrams that are consistent with
the phase diagram of \urusi\ are $\Bog$ and $\Btg$. Of these, only a staggered
$\Bog$ \op\ is consistent with zero-pressure \rus\ data, with a cusp appearing
in the associated elastic modulus. In this picture, the \ho\ phase is
characterized by uniaxial modulated $\Bog$ order, while the high pressure phase
is characterized by uniform $\Bog$ order. The staggered nematic of \ho\ is
similar to the striped superconducting phase found in LBCO and other
cuperates.\cite{Berg_2009b}

We can also connect our results to the large body of work concerning various
multipolar orders as candidate states for \ho\ (e.g.  refs.~\cite{Haule_2009,
Ohkawa_1999, Santini_1994, Kiss_2005, Kung_2015, Kusunose_2011_On}).
Physically, our phenomenological order parameter could correspond to $\Bog$
multipolar ordering originating from the localized component of the U-5f
electrons. For the crystal field states of \urusi, this could correspond either
to electric quadropolar or hexadecapolar order based on the available
multipolar operators. \cite{Kusunose_2011_On} 

The coincidence of our theory's orthorhombic high-pressure phase and \urusi's
\afm\ is compelling, but our mean field theory does not make any explicit
connection with the physics of \afm. Neglecting this physics could be
reasonable since correlations often lead to \afm\ as a secondary effect, like
what occurs in many Mott insulators. An electronic theory of this phase diagram
may find that the \afm\ observed in \urusi\ indeed follows along with an
independent high-pressure orthorhombic phase associated with uniform $\Bog$
electronic order. 

The corresponding prediction of uniform $\Bog$ symmetry breaking in the high
pressure phase is consistent with recent diffraction experiments,
\cite{Choi_2018} except for the apparent earlier onset in temperature of the
$\Bog$ symmetry breaking, which we believe may be due to fluctuating order at
temperatures above the actual transition temperature.  This work motivates both
further theoretical work regarding a microscopic theory with modulated $\Bog$
order, and preforming symmetry-sensitive thermodynamic experiments at pressure,
such as pulse-echo ultrasound, that could further support or falsify this idea.

\begin{acknowledgements}
  Jaron Kent-Dobias is supported by NSF DMR-1719490, Michael Matty is supported
  by NSF DMR-1719875, and Brad Ramshaw is supported by NSF DMR-1752784. We are
  grateful for helpful discussions with Sri Raghu, Steve Kivelson, Danilo
  Liarte, and Jim Sethna, and for permission to reproduce experimental data in
  our figure by Elena Hassinger. We thank Sayak Ghosh for \rus\ data.
\end{acknowledgements}

\appendix

\section{Adding a higher-order interaction}
\label{sec:higher-order}

In this appendix, we compute the $\Bog$ modulus for a theory with a high-order
interaction truncation to better match the low-temperature behavior.  Consider
the free energy density $f=f_\ee+f_\ii+f_\op$ with
\begin{equation}
  \begin{aligned}
    f_\ee&=\frac12C_0\epsilon^2 \\
    f_\ii&=-b\epsilon\eta+\frac12g\epsilon^2\eta^2 \\
    f_\op&=\frac12\big[r\eta^2+c_\parallel(\nabla_\parallel\eta)^2+c_\perp(\nabla_\perp\eta)^2+D(\nabla_\perp^2\eta)^2\big]+u\eta^4.
  \end{aligned}
  \label{eq:new_free_energy}
\end{equation}
The mean-field stain conditioned on the order parameter is found from
\begin{equation}
  \begin{aligned}
    0
    &=\frac{\delta F[\eta,\epsilon]}{\delta\epsilon(x)}\bigg|_{\epsilon=\epsilon_\star[\eta]} \\
    &=C_0\epsilon_\star[\eta](x)-b\eta(x)+g\epsilon_\star[\eta](x)\eta(x)^2,
  \end{aligned}
\end{equation}
which yields
\begin{equation}
  \epsilon_\star[\eta](x)=\frac{b\eta(x)}{C_0+g\eta(x)^2}.
  \label{eq:epsilon_star}
\end{equation}
Upon substitution into \eqref{eq:new_free_energy} and expanded to fourth order
in $\eta$, $F[\eta,\epsilon_\star[\eta]]$ can be written in the form
$F_\op[\eta]$ alone with $r\to\tilde r=r-b^2/C_0$ and $u\to\tilde
u=u+b^2g/2C_0^2$. The phase diagram in $\eta$ follows as before with the
shifted coefficients, and namely $\langle\eta(x)\rangle=\eta_*\cos(q_*x_3)$ for
$\tilde r<c_\perp^2/4D=\tilde r_c$ with $q_*^2=-c_\perp/2D$ and
\begin{equation}
  \eta_*^2=\frac{c_\perp^2-4D\tilde r}{12D\tilde u}
  =\frac{|\Delta\tilde r|}{3\tilde u}.
\end{equation}
We would like to calculate the $q$-dependent modulus
\begin{equation}
  C(q)
  =\frac1V\int dx\,dx'\,C(x,x')e^{-iq(x-x')},
\end{equation}
where
\begin{widetext}
\begin{equation}
  C(x,x')
  =\frac{\delta^2F[\eta_\star[\epsilon],\epsilon]}{\delta\epsilon(x)\delta\epsilon(x')}\bigg|_{\epsilon=\langle\epsilon\rangle}
    =\frac{\delta^2F_\ee[\eta_\star[\epsilon],\epsilon]}{\delta\epsilon(x)\delta\epsilon(x')}+
     \frac{\delta^2F_\ii[\eta_\star[\epsilon],\epsilon]}{\delta\epsilon(x)\delta\epsilon(x')}+
     \frac{\delta^2F_\op[\eta_\star[\epsilon],\epsilon]}{\delta\epsilon(x)\delta\epsilon(x')}
     \bigg|_{\epsilon=\langle\epsilon\rangle}
\end{equation}
and $\eta_\star$ is the mean-field order parameter conditioned on the strain defined implicitly by
\begin{equation}
  0=\frac{\delta F[\eta,\epsilon]}{\delta\eta(x)}\bigg|_{\eta=\eta_\star[\epsilon]}
  =-b\epsilon(x)+g\epsilon(x)^2\eta_\star[\epsilon](x)+\frac{\delta F_\op[\eta]}{\delta\eta(x)}\bigg|_{\eta=\eta_\star[\epsilon]}.
  \label{eq:eta_star}
\end{equation}
We will work this out term by term. The elastic term is the most straightforward, giving
\begin{equation}
  \frac{\delta^2F_\ee[\epsilon]}{\delta\epsilon(x)\delta\epsilon(x')}
  =\frac12C_0\frac{\delta^2}{\delta\epsilon(x)\delta\epsilon(x')}\int dx''\,\epsilon(x'')^2
  =C_0\delta(x-x').
\end{equation}
The interaction term gives
\begin{equation}
  \begin{aligned}
    \frac{\delta^2F_\ii[\eta_\star[\epsilon],\epsilon]}{\delta\epsilon(x)\delta\epsilon(x')}
    &=-b\frac{\delta^2}{\delta\epsilon(x)\delta\epsilon(x')}\int dx''\,\epsilon(x'')\eta_\star[\epsilon](x'')
      +\frac12g\frac{\delta^2}{\delta\epsilon(x)\delta\epsilon(x')}\int dx''\,\epsilon(x'')^2\eta_\star[\epsilon](x'')^2 \\
    &=-b\frac{\delta\eta_\star[\epsilon](x')}{\delta\epsilon(x)}
    -b\frac{\delta}{\delta\epsilon(x)}\int dx''\,\epsilon(x'')\frac{\delta\eta_\star[\epsilon](x'')}{\delta\epsilon(x')}
    +g\frac{\delta}{\delta\epsilon(x)}\big[\epsilon(x')\eta_\star[\epsilon](x')^2\big] \\
    &\qquad+g\frac{\delta}{\delta\epsilon(x)}\int dx''\,\epsilon(x'')^2\eta_\star[\epsilon](x'')\frac{\delta\eta_\star[\epsilon](x'')}{\delta\epsilon(x')} \\
    &=-2(b-2g\epsilon(x)\eta_\star[\epsilon](x))\frac{\delta\eta_\star[\epsilon](x)}{\delta\epsilon(x')}-b\int dx''\,\epsilon(x'')\frac{\delta^2\eta_\star[\epsilon](x'')}{\delta\epsilon(x)\delta\epsilon(x')}
    +g\eta_\star[\epsilon](x)^2\delta(x-x') \\
    &\qquad+g\int dx''\,\epsilon(x'')^2\frac{\delta\eta_\star[\epsilon](x'')}{\delta\epsilon(x)}\frac{\delta\eta_\star[\epsilon](x'')}{\delta\epsilon(x')}
    +g\int dx''\,\epsilon(x'')^2\eta_\star[\epsilon](x'')\frac{\delta^2\eta_\star[\epsilon](x'')}{\delta\epsilon(x)\delta\epsilon(x')}.
  \end{aligned}
\end{equation}
The order parameter term relies on some other identities. First, \eqref{eq:eta_star} implies
\begin{equation}
  \frac{\delta F_\op[\eta]}{\delta\eta(x)}\bigg|_{\eta=\eta_\star[\epsilon]}
    =b\epsilon(x)-g\epsilon(x)^2\eta_\star[\epsilon](x),
  \label{eq:dFodeta}
\end{equation}
and therefore that the functional inverse $\eta_\star^{-1}[\eta]$ is
\begin{equation}
  \eta_\star^{-1}[\eta](x)=\frac{b}{2g\eta(x)}\Bigg(1-\sqrt{1-\frac{4g\eta(x)}{b^2}\frac{\delta F_\op[\eta]}{\delta\eta(x)}}\Bigg).
\end{equation}
The inverse function theorem further implies (with substitution of \eqref{eq:dFodeta} after the derivative is evaluated) that
\begin{equation}
  \bigg(\frac{\delta\eta_\star[\epsilon](x)}{\delta\epsilon(x')}\bigg)^{\{-1\}}
  =\frac{\delta\eta_\star^{-1}[\eta](x)}{\delta\eta(x')}\bigg|_{\eta=\eta_\star[\epsilon]}
  =\frac{g\epsilon(x)^2\delta(x-x')+\frac{\delta^2F_\op[\eta]}{\delta\eta(x)\delta\eta(x')}\big|_{\eta=\eta_\star[\epsilon]}}{b-2g\epsilon(x)\eta_\star[\epsilon](x)}
\end{equation}
and therefore that
\begin{equation}
  \frac{\delta^2F_\op[\eta]}{\delta\eta(x)\delta\eta(x')}\bigg|_{\eta=\eta_\star[\epsilon]}
  =(b-2g\epsilon(x)\eta_\star[\epsilon](x))\bigg(\frac{\delta\eta_\star[\epsilon](x)}{\delta\epsilon(x')}\bigg)^{\{-1\}}
  -g\epsilon(x)^2\delta(x-x').
  \label{eq:d2Fodetadeta}
\end{equation}
Finally, we evaluate the order parameter term, using \eqref{eq:dFodeta} and \eqref{eq:d2Fodetadeta} which give
\begin{equation}
  \begin{aligned}
    \frac{\delta^2F_\op[\eta_\star[\epsilon]]}{\delta\epsilon(x)\delta\epsilon(x')}
    &=\frac{\delta}{\delta\epsilon(x)}\int dx''\,\frac{\delta\eta_\star[\epsilon](x'')}{\delta\epsilon(x')}\frac{\delta F_\op[\eta]}{\delta\eta(x'')}\bigg|_{\eta=\eta_\star[\epsilon]} \\
    &=\int dx''\,\frac{\delta^2\eta_\star[\epsilon](x'')}{\delta\epsilon(x)\delta\epsilon(x')}\frac{\delta F_\op[\eta]}{\delta\eta(x'')}\bigg|_{\eta=\eta_\star[\epsilon]}
    +\int dx''dx'''\frac{\delta\eta_\star[\epsilon](x'')}{\delta\epsilon(x)}\frac{\delta\eta_\star[\epsilon](x''')}{\delta\epsilon(x')}\frac{\delta^2F_\op[\eta]}{\delta\eta(x'')\delta\eta(x''')}\bigg|_{\eta=\eta_\star[\epsilon]} \\
    &=\int dx''\,\frac{\delta^2\eta_\star[\epsilon](x'')}{\delta\epsilon(x)\delta\epsilon(x')}\big(b\epsilon(x)-g\epsilon(x)^2\eta_\star[\epsilon](x)\big)
    +(b-2g\epsilon(x)\eta_\star[\epsilon](x))\frac{\delta\eta_\star[\epsilon](x)}{\delta\epsilon(x')} \\
    &\qquad-g\int dx''\,\epsilon(x'')^2\frac{\delta\eta_\star[\epsilon](x'')}{\delta\epsilon(x)}\frac{\delta\eta_\star[\epsilon](x'')}{\delta\epsilon(x')}.
  \end{aligned}
\end{equation}
Summing all three terms, we see a great deal of cancellation, with
\[
  \frac{\delta^2F[\eta_\star[\epsilon],\epsilon]}{\delta\epsilon(x)\delta\epsilon(x')}=C_0\delta(x-x')+g\eta_\star[\epsilon](x)^2\delta(x-x')-(b-2g\epsilon(x)\eta_\star[\epsilon](x))\frac{\delta\eta_\star[\epsilon](x)}{\delta\epsilon(x')}.
\]
We new need to evaluate this at $\langle\epsilon\rangle$. First, $\eta_\star[\langle\epsilon\rangle]=\langle\eta\rangle$, and
\[
  \frac{\delta^2F[\eta_\star[\epsilon],\epsilon]}{\delta\epsilon(x)\delta\epsilon(x')}\bigg|_{\epsilon=\langle\epsilon\rangle}=C_0\delta(x-x')+g\langle\eta(x)\rangle^2\delta(x-x')-(b-2g\langle\epsilon(x)\rangle\langle\eta(x)\rangle)\frac{\delta\eta_\star[\epsilon](x)}{\delta\epsilon(x')}\bigg|_{\epsilon=\langle\epsilon\rangle}.
\]
Computing the final functional derivative is the most challenging part. We will
first compute its functional inverse, take the Fourier transform of that, and
then use the basic relationship between Fourier functional inverses to find the
form of the non-inverse. First, we note
\begin{equation}
  \frac{\delta^2F_\op[\eta]}{\delta\eta(x)\delta\eta(x')}\bigg|_{\eta=\langle\eta\rangle}
  =\big[r-c_\perp\nabla_\perp^2-c_\parallel\nabla_\parallel^2+D\nabla_\perp^4+12u\langle\eta(x)\rangle^2\big]\delta(x-x'),
\end{equation}
which gives
\begin{equation}
  \begin{aligned}
    \bigg(\frac{\delta\eta_\star[\epsilon](x)}{\delta\epsilon(x')}\bigg)^{\{-1\}}\bigg|_{\epsilon=\langle\epsilon\rangle}
    &=\frac1{b-2g\langle\epsilon(x)\rangle\langle\eta(x)\rangle}\bigg[g\langle\epsilon(x)\rangle^2\delta(x-x')+\frac{\delta^2F_\op[\eta]}{\delta\eta(x)\delta\eta(x')}\bigg]_{\eta=\langle\eta\rangle} \\
    &=\frac1{b-2g\langle\epsilon(x)\rangle\langle\eta(x)\rangle}\Big[
      g\langle\epsilon(x)\rangle^2+r-c_\perp\nabla_\perp^2-c_\parallel\nabla_\parallel^2+D\nabla_\perp^4+12u\langle\eta(x)\rangle^2
    \Big]\delta(x-x').
  \end{aligned}
\end{equation}
Upon substitution of \eqref{eq:epsilon_star} and expansion to quadratic order it $\langle\eta(x)\rangle$, we find
\begin{equation}
  \begin{aligned}
    \bigg(\frac{\delta\eta_\star[\epsilon](x)}{\delta\epsilon(x')}\bigg)^{\{-1\}}\bigg|_{\epsilon=\langle\epsilon\rangle}
    &=\frac1b\Bigg\{r-c_\perp\nabla_\perp^2-c_\parallel\nabla_\parallel^2+D\nabla_\perp^4\\
    &\qquad\qquad+\langle\eta(x)\rangle^2\bigg[12u+\frac{b^2g}{C_0^2}+\frac{2g}{C_0}(r-c_\perp\nabla_\perp^2-c_\parallel\nabla_\parallel^2+D\nabla_\perp^4)\bigg]+O(\langle\eta\rangle^4)\Bigg\}\delta(x-x').
  \end{aligned}
\end{equation}
Defining $\widehat{\langle\eta\rangle^2}=\int dq'\,\langle\hat\eta(q')\rangle\langle\hat\eta(-q')\rangle$, its Fourier transform is then
\begin{equation}
  \begin{aligned}
    G(q)
    &=\frac1V\int dx\,dx'\,e^{-iq(x-x')}\bigg(\frac{\delta\eta_\star[\epsilon](x)}{\delta\epsilon(x')}\bigg)^{\{-1\}}\bigg|_{\epsilon=\langle\epsilon\rangle} \\
    &=\frac1b\Bigg\{r+c_\perp q_\perp^2+c_\parallel q_\parallel^2+Dq_\perp^4+\widehat{\langle\eta\rangle^2}\bigg[12u+\frac{b^2g}{C_0^2}+\frac{2g}{C_0}(r+c_\perp q_\perp^2+c_\parallel q_\parallel^2+Dq_\perp^4)\bigg]+O(\langle\hat\eta\rangle^4)\Bigg\}.
  \end{aligned}
\end{equation}
We can now compute $C(q)$ by taking its Fourier transform, using the convolution theorem for the second term:
\begin{equation}
  \begin{aligned}
    C(q)
    &=C_0+g\widehat{\langle\eta\rangle^2}-\int dq''\bigg(b\delta(q'')-\frac{gb}{C_0}\int dq'\langle\hat\eta_{q'}\rangle\langle\hat\eta_{q''-q'}\rangle\bigg)/G(q-q'') \\
    &=C_0+g\widehat{\langle\eta\rangle^2}-b^2\bigg(\frac1{r+c_\perp q_\perp^2+c_\parallel q_\parallel^2+Dq_\perp^4}-\widehat{\langle\eta\rangle^2}\frac{12u+b^2g/C_0^2+\frac{2g}{C_0}(r+c_\perp q^2+c_\parallel q_\parallel^2+Dq_\perp^4)}{(r+c_\perp q_\perp^2+c_\parallel q_\parallel^2+Dq_\perp^4)^2}\bigg)\\
    &\qquad+\frac{gb^2}{C_0}\int dq'\,dq''\frac{\langle\hat\eta_{q'}\rangle\langle\hat\eta_{q''-q'}\rangle}{r+c_\perp(q_\perp-q_\perp'')^2+c_\parallel(q_\parallel-q_\parallel'')^2+D(q_\perp-q_\perp'')^4}+O(\langle\hat\eta\rangle^4).
  \end{aligned}
\end{equation}
Upon substitution of $\langle\hat\eta_q\rangle=\frac12\eta_*\big[\delta(q_\perp-q_*)+\delta(q_\perp+q_*)\big]\delta(q_\parallel)$, we have
\begin{equation}
  \begin{aligned}
    C(q)
    &=C_0+\frac14g\eta_*^2-b^2\bigg(\frac1{r+c_\perp q_\perp^2+c_\parallel q_\parallel^2+Dq_\perp^4}-\frac{\eta_*^2}4\frac{12u+b^2g/C_0^2+\frac{2g}{C_0}(r+c_\perp q^2+c_\parallel q_\parallel^2+Dq_\perp^4)}{(r+c_\perp q_\perp^2+c_\parallel q_\parallel^2+Dq_\perp^4)^2}\bigg)\\
    &\qquad+\frac{gb^2\eta_*^2}{4C_0}\bigg(\frac2{r+c_\parallel q_\parallel^2+c_\perp q_\perp^2+Dq_\perp^4}+\frac1{r+c_\parallel q_\parallel^2+c_\perp(q_\perp-2q_*)^2+D(q_\perp-2q_*)^4} \\
    &\qquad\qquad+\frac1{r+c_\parallel q_\parallel^2+c_\perp(q_\perp+2q_*)^2+D(q_\perp+2q_*)^4}\bigg)+O(\eta_*^4).
  \end{aligned}
\end{equation}
Evaluating at $q=0$, we have
\begin{equation}
  \begin{aligned}
    C(0)
    &=C_0-\frac{b^2}r+\frac{\eta_*^2}4\bigg(g+\frac{b^2}{r^2}(12u+b^2g/C_0^2)+\frac{2gb^2}{C_0r}\frac{16Dq_*^4+3r}{8Dq_*^4+r}\bigg)
  \end{aligned}
  \label{eq:C0}
\end{equation}
\end{widetext}
Above the transition this has exactly the form of \eqref{eq:static_modulus} for
any $g$; below the transition it has the same form at $g=0$ to order
$\eta_*^2$. With $r=a\Delta T+c^2/4D+b^2/C_0$, $u=\tilde u-b^2g/2C_0^2$, and
\begin{equation}
  \eta_*^2=\begin{cases}
    0 & \Delta T > 0 \\
    -a\Delta T/3\tilde u & \Delta T \leq 0,
  \end{cases}
\end{equation}
we can fit the ratios $b^2/a=1665\,\mathrm{GPa}\,\mathrm K$, $b^2/Dq_*^4=6.28\,\mathrm{GPa}$, and $b\sqrt{-g/\tilde u}=14.58\,\mathrm{GPa}$ with $C_0=(71.14-(0.010426\,\mathrm K^{-1})T)\,\mathrm{GPa}$. The resulting fit is shown as a dashed black line in Fig.~\ref{fig:data}.

\bibliographystyle{apsrev4-1}
\bibliography{hidden_order}

\end{document}